\documentclass[runningheads,a4paper]{llncs}

\usepackage{amssymb}
\usepackage{graphicx}
\usepackage{epstopdf}

\usepackage{url}
\urldef{\mailsa}\path|{rrychtarikova, stys}@frov.jcu.cz|    
\newcommand{\keywords}[1]{\par\addvspace\baselineskip
\noindent\keywordname\enspace\ignorespaces#1}

\begin{document}

\mainmatter  

\title{Information limits of optical microscopy: application to fluorescently labelled tissue section}

\titlerunning{Information limits of optical microscopy}

\author{Renata Rycht\'{a}rikov\'{a}\inst{1} \and Georg Steiner\inst{2} \and Michael B. Fischer\inst{3} \and Dalibor \v{S}tys\inst{1}}

\authorrunning{Rycht\'{a}rikov\'{a} et al.}

\institute{University of South Bohemia in \v{C}esk\'{e} Bud\v{e}jovice, Faculty of Fisheries and Protection of Waters, South Bohemian Research Center of Aquaculture and Biodiversity of Hydrocenoses, Institute of Complex Systems, Z\'{a}mek 136, 373 33 Nov\'{e} Hrady, Czech Republic\\
\mailsa\\
\url{http://www.frov.jcu.cz/cs/ustav-komplexnich-systemu-uks}
\and
TissueGnostics GmbH, Taborstrasse 10/2/8, 1020 Vienna, Austria
\and
Danube University Krems, Department of Health Sciences and Biomedicine, Dr.-Karl-Dorrek-Strasse 30, 3500 Krems, Austria}

\toctitle{Lecture Notes in Computer Science}
\tocauthor{Rycht\'{a}rikov\'{a} et al.}
\maketitle

\begin{abstract}
The article demonstrates some less known principles of image build-up in diffractive microscopy and their usage in analysis unravelling the smallest localized information about the original object -- an electromagnetic centroid. In fluorescence, the electromagnetic centroid is naturally at the position of the fluorophore. The usage of an information-entropic variable -- a point divergence gain -- is demonstrated for finding the most localized position of the object's representation, generally of the size of a voxel (3D pixel). These spatial pixels can be qualitatively classified and used for reconstruction of the 3D structures with precision comparable with electron microscopy.
\end{abstract}

\section{Introduction}

In optical microscopy, features of a light-intensity profile arise essentially by an interplay of two phenomena: by interaction of light with the matter and by modification of the light path by the microscope~\cite{Thorn2016}. In observation of living cells there are mainly two types of responses of the interaction of light with the matter: (i) diffraction of light on particles and (ii) emission of fluorescent light after absorption. 

The diffraction of light is a process whose proper description is difficult, in most real cases essentially impossible~\cite{Mie1909}. A general feature of light behaviour behind a scattering object is the existence of a dark "spire" which is gradually narrowing as the Huygens waves cover the space. The light diffraction on an object which is circular enough can even produce bright spots on axis, so-called the Arago spots~\cite{Fresnel1868}. Biological objects are in most cases formed by dense structured assemblies of proteins, nucleic acids, lipids and other molecules. Objects in the cell interior observable by light diffraction are not merely objects of in-average different refractivity index but they are also internally inhomogeneous. Their detailed scattering properties are beyond a simple description. We are thus content with the information provided by the microscopy experiment, although we have to understand the limitations of this information. 

Light fluorescence has been an important examination tool in biological cell for a long time~\cite{Thorn2016}. The fluorescent microscopy of a living cell is in the forefront of the interest due to the efforts to achieve/break the limits of resolution of the observed intracellular objects~\cite{nobelprize2016}. All attempts at superresolution were theoretically based on the description of observed phenomena using the Maxwell theory of the electromagnetic field which is a proper description of behaviour of an ensemble of photons~\cite{Maxwell}. 

We have recently shown~\cite{Rychtarikovaetal2016} that information projected by light on the screen of a camera can be analyzed down to the level of a single sensor element of the digital camera or, in other words, of a single pixel in a digital image. A single pixel whose size can be determined can be referred to the location of an observed object in the original image. The size of this element does not have any physical limit, i.e., it can be of a-few-nanometer size. This finding contradicts no principles of quantum mechanics, because we do not determine the position of a single photon but the most probable location of a large ensemble of photons whose distribution have a maximum, intensity profile, etc. 

Finding the maximum or minimum of an intensity profile does not mean that the object, which gives rise to the response, is located at the particular position in space to which a microscope image seemingly refers. In diffractive imaging, the reason is that both the darkest and smallest point is located outside the object which causes the diffraction. The maximum of fluorescence of a fluorescent object, e.g. a bead, is (in the ideal case) seemingly on the surface at a place closest to the surface of the objective lens. The exact maximum or minimum can be defined as a centroid of the outcome of the electromagnetic process which gives rise to the observed light phenomenon. 

The attempt to yield a maximal information from the digital image exposes also all non-idealities of the optical path. In case that the resulted image is spectrally resolved, i.e. when a colour digital camera is used, each camera channel typically detects different information~\cite{Rychtarikovaetal2016,Stysetal2016,Cisaretal2016}. The difference is generally attributed to the composition of the object which gives rise to the signal, which is ultimately true, but the exact way how this difference is conveyed by the optical path is not easy to unravel. Namely, modern apochromatic lenses utilize combinations of lenses to project all colours at the same place. This assumption is indeed valid only with a finite precision and can be challenged when minute image details are interpreted. 

The microscope observation should answer these questions:

\begin{enumerate}
\item Where is located the object giving rise to the response?
\item What is the shape of the object? 
\item What are the spectral characteristics of the object?
\end{enumerate}

In order to answer these questions, under the microscope, we have analyzed a response of a standard object -- a single microparticle -- and a section of fluorescently labelled tissue. We introduce a method which systematically determines the electromagnetic centroid of a diffracting object as a centroid of the information in 3D space with the precision of a single voxel (3D pixel).

\section{Materials and methods}
\label{methods}

\subsection{Experiment on latex particle}

A latex particle of the diameter of 2000 nm was placed on a carbon layer on a electron microscopy copper grid covered by amorphous carbon (prepared at the Institute of Parasitology AS CR, \v{C}esk\'{e} Bud\v{e}jovice, CZ). The sample was scanned under a optical transmission microscope~\cite{Rychtarikovaetal2016} -- nanoscope (Institute of Complex Systems, Nov\'{e} Hrady, CZ) -- equipped by a 12-bit colour Kodak KAI-16000 digital camera with a chip of 4872$\times$3248 resolution (Camera Offset 200, Camera Gain 383, Camera Exposure 2950 ms). A Nikon objective (60$\times$/0.8, $\infty$/0.17, WD 0.3) which gives the resulted size of the image pixel of 46$\times$46 nm$^2$ was used. The sample was illuminated by two Luminus 360 LEDs charged by the current of 4000 mA. The standard deviation of the z-step was minimized by the pngparser.exe software~\cite{Rychtarikovaetal2016} which gave a z-stack of 258 images of the average step of 152 nm. 

\subsection{Experiment on prostate cancer tissue}

A fixed sample of prostate cancer tissue was imunnofluorescently labelled by DAPI (4',6-Diamidine-2'-phenylindole) and Anti-Cytokeratin 18. The treated sample was scanned using a Axio Imager Z2 fluorescent microscope (TissueGnostics GmbH, Vienna, AT) equipped by a 12-bit grayscale Andor camera Zyla 5.5 with a chip of 1560$\times$1960 resolution. A 100$\times$ oil objective (1.3 NA) gave the resulted image pixel of the size of 328$\times$328 nm$^2$. The microscope step along the z-axis was 100 nm. The full z-stack series contained 82 images.

\subsection{Image processing and visualization}

The 3D reconstructions of the image z-stacks were performed by the method similar to that described in detail in~\cite{Rychtarikovaetal2016}. The modifications were as follows:
\begin{enumerate}
\item The segmentation of the object of interest using an algorithm which searches for intensities of unchanged values between two consecutive images was applied only to the z-stack of the bead, not to the z-stack of the dyed tissue.    
\item The main difference was in the selection of a focused part of both z-stacks using k-means clustering into two groups using Unscrambler$^{\textregistered}$ X software, Norway. Instead of vectors of point information gain, vectors of point divergence gain~\cite{Rychtarikovaetal2015} were computed from intensity histograms of the whole images for a set of $\alpha$ = \{0.1, 0.3, 0.5, 0.7, 0.99, 1.3, 1.5, 1.7, 2.0, 2.5, 3.0, 3.5, 4.0 (the bead), 5.0, 6.0, 7.0 (the tissue)\}. In case of the 2-$\mu$m bead scanned in light transmission, the z-stack which underwent the computation of the point divergence gain vectors, was firstly transformed into 8 bits by the Least Information Lost (LIL) algorithms~\cite{Stysetal2016} which enables to yield the maximum of information during the bit-depth reduction and to compare images through the whole stack. The clustering reduced the number of images from 258 to 81 (Imgs. 83--163) and from 82 to 52 (Imgs. 26--77) for the bead and the tissue, respectively.  
\item The points of almost unchanged intensities in two consecutive in-focus images were stacked after (i) segmentation of the object demarcated by green intensity 1000 in the first detectable image followed by the calculation of point divergence gain for $\alpha=4.0$ (the bead) and (ii) computation of the point divergence gain for $\alpha$ of the value of 7.0 and 6.0 for DAPI and Anti-Cytokeratin 18, respectively, followed by the automatic selection of two kinds of relevant points as written in Section~\ref{cells}. 
\end{enumerate}

The figures of sections of the point spread function of the microbead in each colour channel as well as of positions of the electromagnetic centroids (Fig.~\ref{fig:real_psf}) were plotted using Matlab$^{\textregistered}$ 2016b (Mathworks, USA) software.  

All microscopic figures in the article were also visualized using the LIL conversion into 8 bits. The original and processed image sets and relevant Matlab algorithms are available at~\cite{ftp}.

\section{Results}

\subsection{Electromagnetic centroid}

\begin{figure}[h]
\centering
\includegraphics[height=5cm]{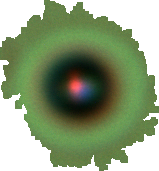}
\caption{Image 127 from a z-stack of microscopic images of a 2000-nm latex particle.} 
\label{fig:img_127}
\end{figure}

Fig.~\ref{fig:img_127} shows a typical example of a microscopic image of the simplest object -- a latex particle of the diameter of 2 $\mu$m -- at the position of visually determined focus. For each (red, green, blue) image channel, the response  of the object differs in position and in shape of the intensity distribution.

As a consequence of light interactions, one can find two smallest objects on the apparent axis of the response of interactions of the electromagnetic field with a diffracting object in each colour channel: A dark spot, which is an outcome of the destructive light diffraction, and a bright spot, which is surrounded by dark intensities and is called the Arago spot (Fig.~\ref{fig:img_127}). In the ideal case, this bright spot appears due to constructive light interference at the optical axis at the position of the center of a circular diffracting object~\cite{Fresnel1868}.

\begin{figure}[h]
\centering
\includegraphics[width=.8\textwidth]{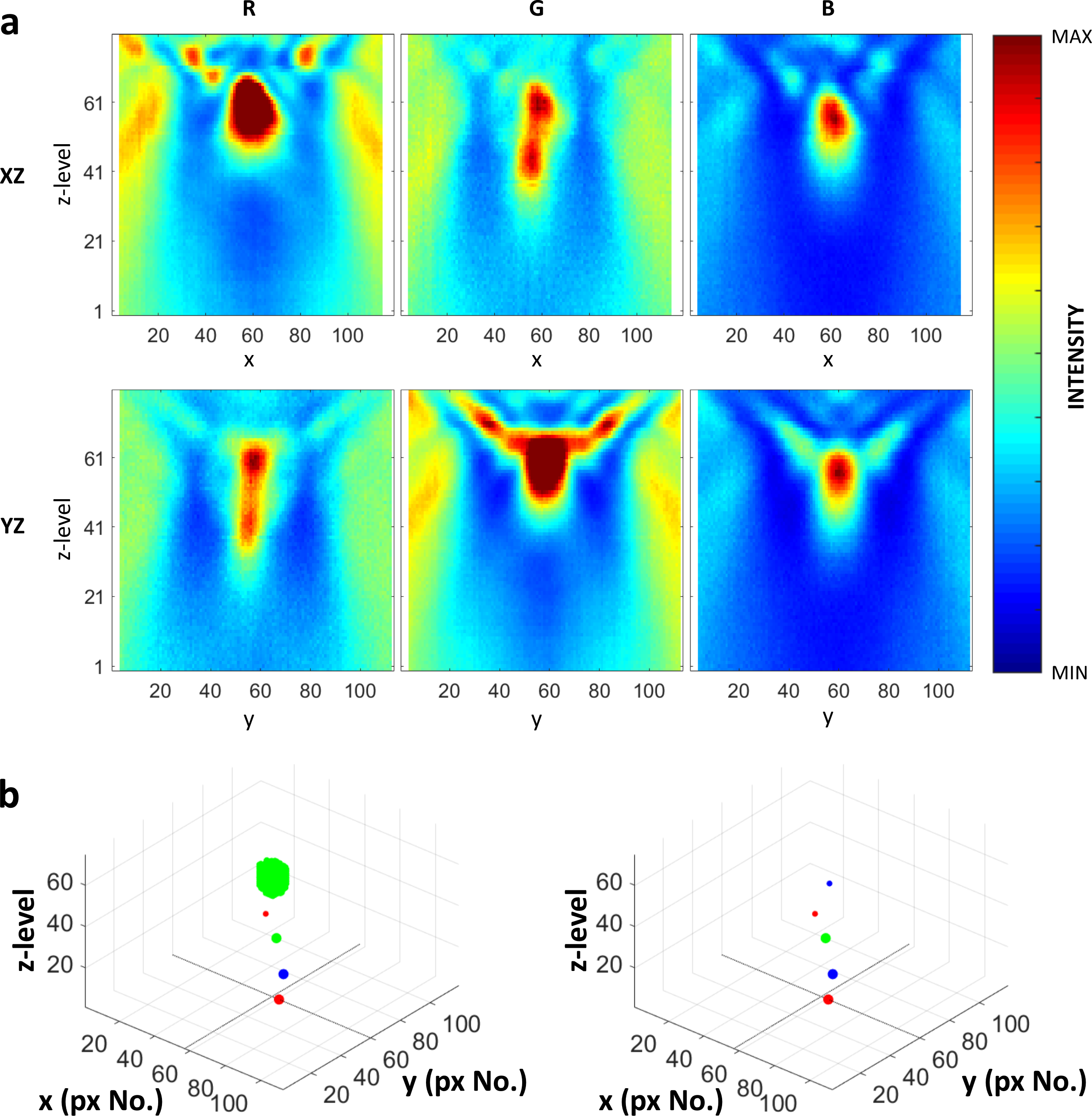}
\caption{3D intensity maps of the 2000-nm latex bead in Fig.~\ref{fig:img_127} in the red, green, and blue image channel. The minimal/maximal intensities are 623/2706 (R), 759/4095 (G), 430/3752 (B). Voxel size is 46 nm (horizonally) and 152 nm (vertically). \textbf{a)} Sections in the xz- and yz-plane, respectively. \textbf{b)} Positions of electromagnetic centroids  (\textit{colour-coded}). The bigger points and the smaller points correspond to negative and positive light interferences, respectively. The bright spot in the green channel is not sufficiently resolved due to the saturation of the signal. The positions of xz- and yz-planes relevant to \textbf{a} are highlighted by bold lines.}
\label{fig:real_psf}
\end{figure}

Fig.~\ref{fig:real_psf}\textbf{a} demonstrates xz- and yz-planes of the 3D microscopic image of the latex particle in~Fig.~\ref{fig:img_127} including all non-idealities of the microscope optics. The detected intensities are combinations of an interaction of the light electromagnetic field with the sample followed by a transformation of the intensities by the microscope optics. Fig.~\ref{fig:real_psf}\textbf{b} depicts the positions of the dark and bright maxima -- the centroids of the electromagnetic field -- in each colour channel. The bright green maximum (Fig.~\ref{fig:real_psf}\textbf{b}, \textit{left}) consists of many points of identical intensities due to the saturation of the 12-bit intensity signal at high intensities.

\subsection{Fluorescently labelled cells}
\label{cells}

\begin{figure}
\centering
\includegraphics[width=0.8\textwidth]{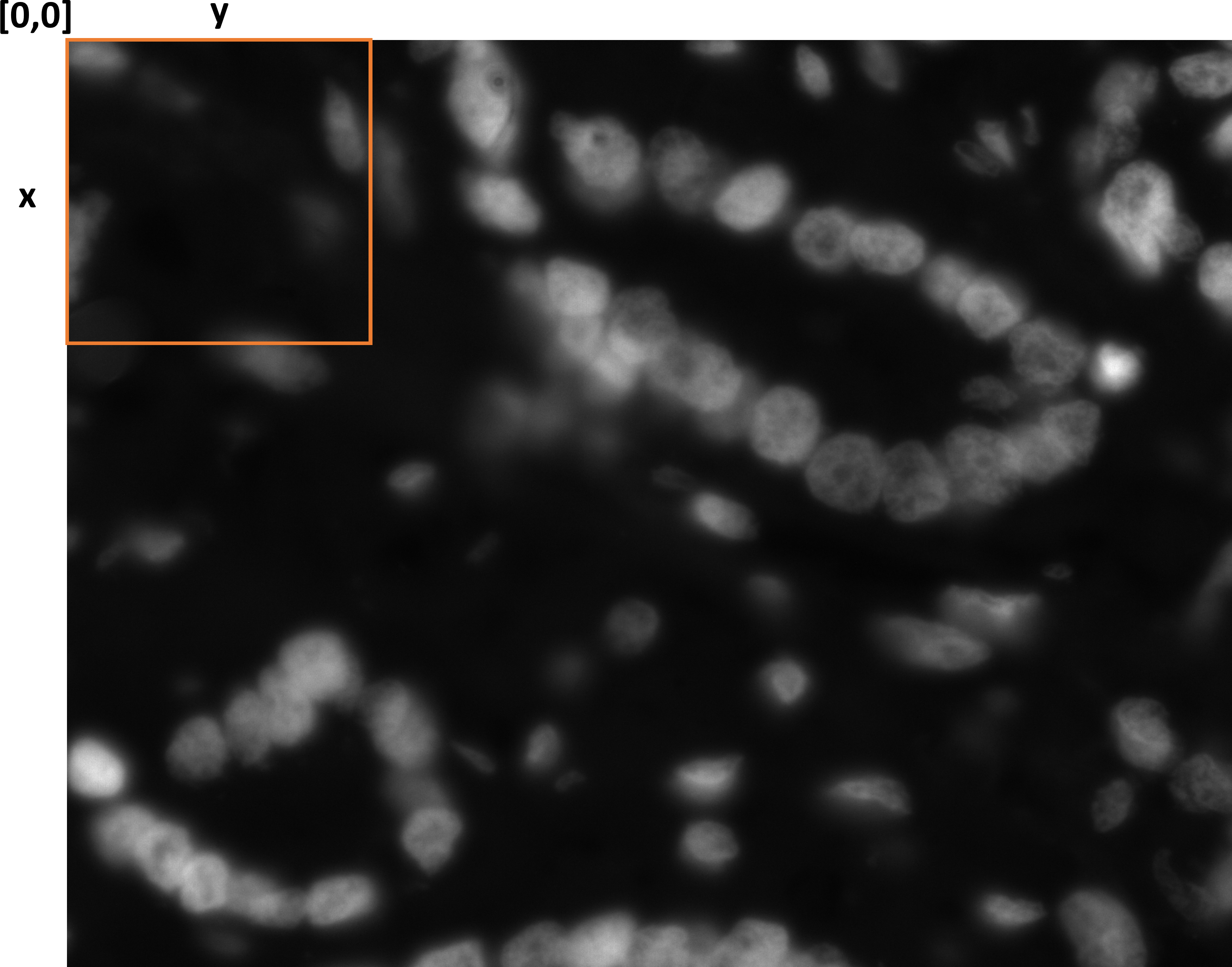}
\caption{Microscopic image of the section of prostate cancer tissue labelled by DAPI dye targeted to nuclei. Pixel size is 328$\times$328 nm$^2$.} 
\label{fig:fluorescent_cells}
\end{figure}

\begin{figure}
\centering
\includegraphics[width=0.8\textwidth]{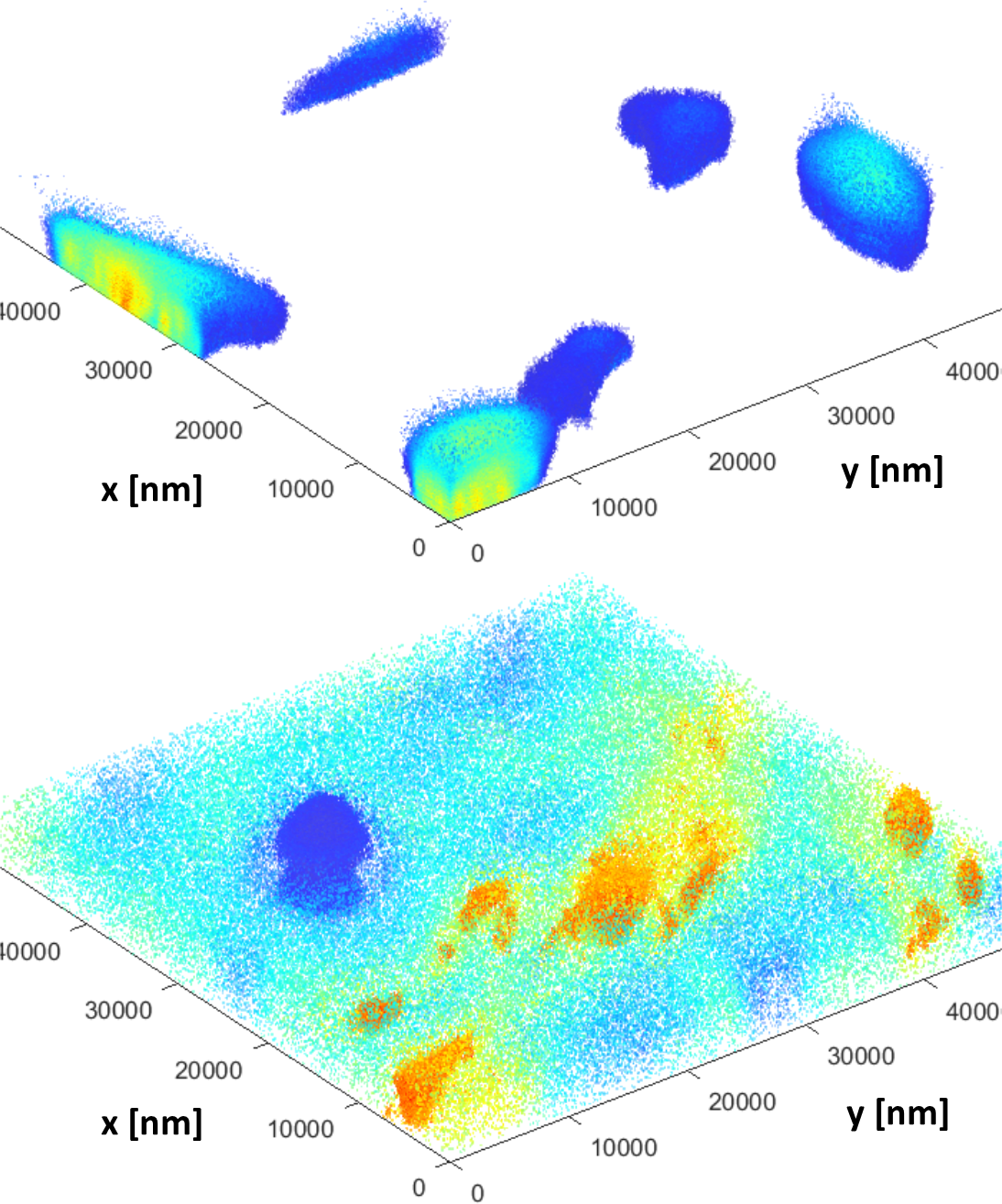}
\caption{3D reconstruction of the positions of points of unchanged intensities in the section of prostate cancer tissue labelled by DAPI dye targeted to nuclei (\textit{upper}) and Anti-Cytokeratin 18 targeted to keratin (\textit{lower}). The section is located in the upper left corner of Fig.~\ref{fig:fluorescent_cells} (\textit{orange section}). Voxel size is 328$\times$328$\times$100 nm$^3$. Colorbar is the same as in Fig.~\ref{fig:real_psf}\textbf{a} but with the minimal/maximal intensities of 684/3495 for DAPI and 248/1522 for Anti-Cytokeratin 18.} 
\label{fig:3D_fluorescent_cells}
\end{figure}

Fig.~\ref{fig:fluorescent_cells} shows a typical microscopic image of a fluorescently labelled tissue, where the observed intensities correspond to the positions of a fluorophore in the sample. The fluorophore changes its spectral properties and quantum efficiency in response to the environment. The information analysis of the given z-stack of the labelled tissue using the point divergence gain was focused on identification of positions of fluorophores which, according to given rules, differ from other (Fig.~\ref{fig:3D_fluorescent_cells}, \textit{upper}). The given rules are to examine (and display) points whose intensity does not change over two z-levels and which were selected according to
\begin{itemize}
\item their proximity, i.e. dense areas are displayed and 
\item different intensity, i.e. spectrally different object are displayed.
\end{itemize}
In less favourable cases, any simple rule is not available and the image has to be analyzed from the complete dataset. For instance, (Fig. \ref{fig:3D_fluorescent_cells}, \textit{lower}) shows an example of the 3D reconstruction which arose from images in which the unchanged points occurred only very sparsely.

\begin{figure}[h]
\begin{tabular}{c c c}
\centering
\includegraphics[width=0.3\textwidth]{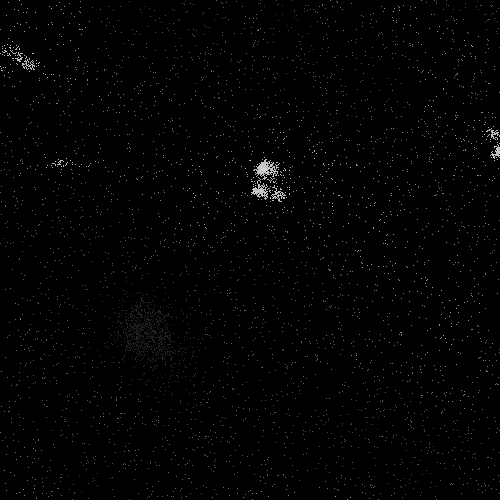} &
\includegraphics[width=0.3\textwidth]{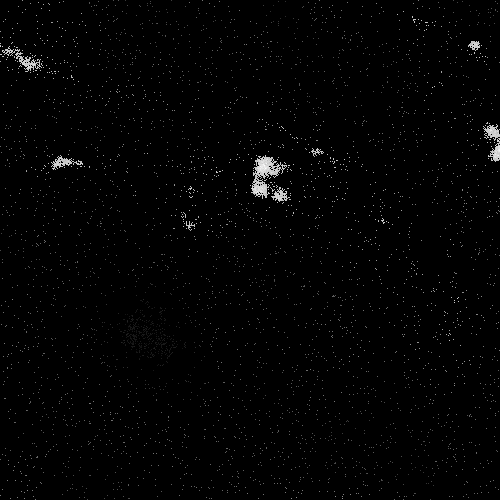} &
\includegraphics[width=0.3\textwidth]{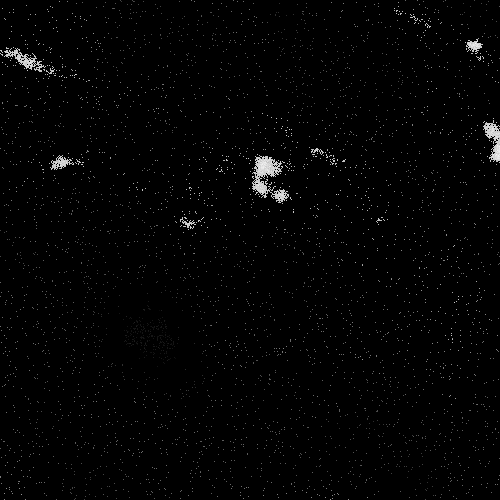} 
\\
\includegraphics[width=0.3\textwidth]{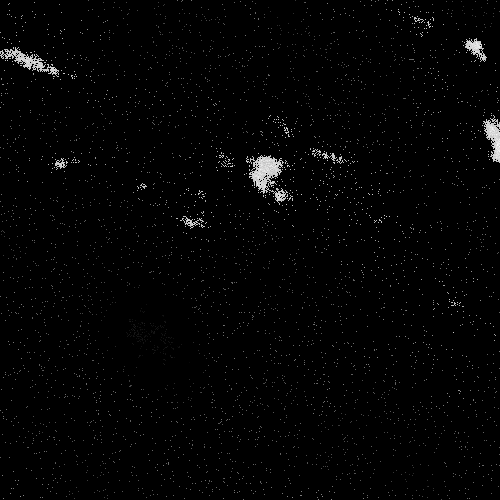} &
\includegraphics[width=0.3\textwidth]{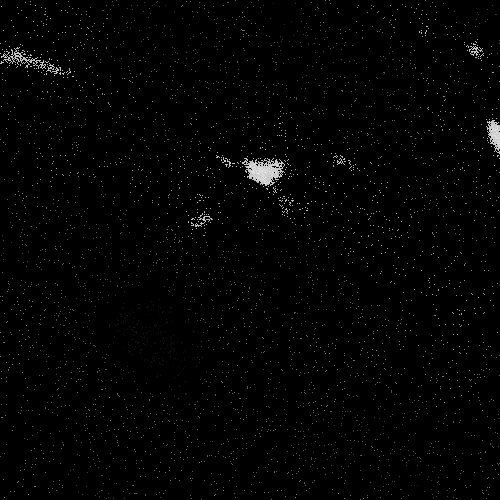} &
\includegraphics[width=0.3\textwidth]{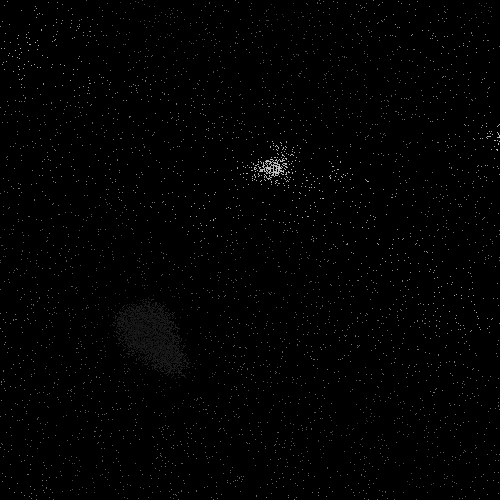}
\end{tabular}
\caption{7$^{\mbox{th}}$, 16$^{\mbox{th}}$, 21$^{\mbox{st}}$, 26$^{\mbox{th}}$, 36$^{\mbox{th}}$, and 46$^{\mbox{th}}$ image (\textit{from upper right to lower left}) of the series from which Fig.~\ref{fig:3D_fluorescent_cells}, \textit(lower), \textit{lower} was constructed. Pixel size is 328$\times$328 nm$^2$.}
\label{fig:original_texa}
\end{figure}

\section{Discussion}

This paper tries to answer how to recognize unchanged information between two optical cuts in optical microscopy in order to find the most localize information about the position of an object. It has few technical determinants:
\begin{enumerate}
\item First of all, the comparison of the optical cuts is limited by the analog-to-digital (AD) conversion. The AD converters in standard digital cameras provide 12- or 16-bit conversion (4096 or  65536 intensity levels). However, experimenters usually work with 8-bit images with 256 intensity levels. Image distortions which accompany the 12/16-bit to 8-bit conversion was reported earlier~\cite{Stysetal2016}. In this paper, we report an analysis of original 12/16-bit datasets. 

\item The second main technical determinant is a size of pixel or voxel (3D pixel) which determines a theoretical size of the object. It is not the goal of this article to discuss limits of the discriminability of objects in terms of the theory of light, we assume hereby that it may be determined experimentally. 

\item The third determinant is the exposure time which is closely related to the sensitivity of camera chip. The exposure of time of camera is usually set so that all intensity signals would be captured without any oversaturation. In addition, the linear responses of the chip elements to exposure time are never guaranteed and the calibration is, according to our experience, never correct.

\item The acquisition time of cameras also plays a pivotal role. In case of observation of moving object, e.g., living cells, the acquisition time can be so long that the object changes its position and a false negative signal is obtained. 

\item Further, the objective determinants of the precision of finding the most localized information about the position of an object come from various sources of noise. In the examples presented in this article, when the samples were observed under high light intensities (i.e. under a large ensemble of photons), it is unlikely that the quantum noise~\cite{Mizushima1988} was the key limit which prevent to find the intensity maximum or minimum. In these examples, the signal distortions originate from numerous sources related to the optical path and the instrument electronics. The noise modulates the signal by the multiplication by a function. Since no precise characteristics of the noise is known, it is most appropriate to assume that it has a multifractal character. This mutifractal intensity response is then discretized in space and in time. A method of calculation of point divergence gain~\cite{Rychtarikovaetal2016} enables to account for the whole spectrum of these distortions. The points (pixels) of almost unchanged intensities are grouped with regards to different assumptions about the distributions of occurrence of intensity signals. We used the method of point divergence gain~\cite{Rychtarikovaetal2016} for resolution of the information in z-stacks of images of living cells into elementary information contributions in the diffractive imaging. In the green image channel, there was observed a lower number of identical points between two consecutive z-levels. It is assigned to a broader intensity spectrum, i.e., a broader wavelength range of the green filters of the Bayer mask~\cite{Bayer}. This illustrates a paradox of the information analysis: The broader the information spectrum is, the more the differences are found. 
\end{enumerate}

A latex bead of the 2-$\mu$m size is an object of the size which is very relevant to intracellular objects such as mitochondria or other oval organelles. Its size in the visible light spectrum is from 5 light waves (shortwave blue, 400 nm) to less than 3 light waves (far red, 720 nm), which is on the border between macroscopic behaviour and behaviour of, e.g., metal nanobeads of the size of the fraction of the light wavelength. If we stick with the macroscopic description, we can say that the Arago spot~\cite{Fresnel1868} is located in the vicinity of the object and is directly adjacent to the dark area of diffraction. 

\begin{figure}
\centering
\includegraphics[width=\textwidth]{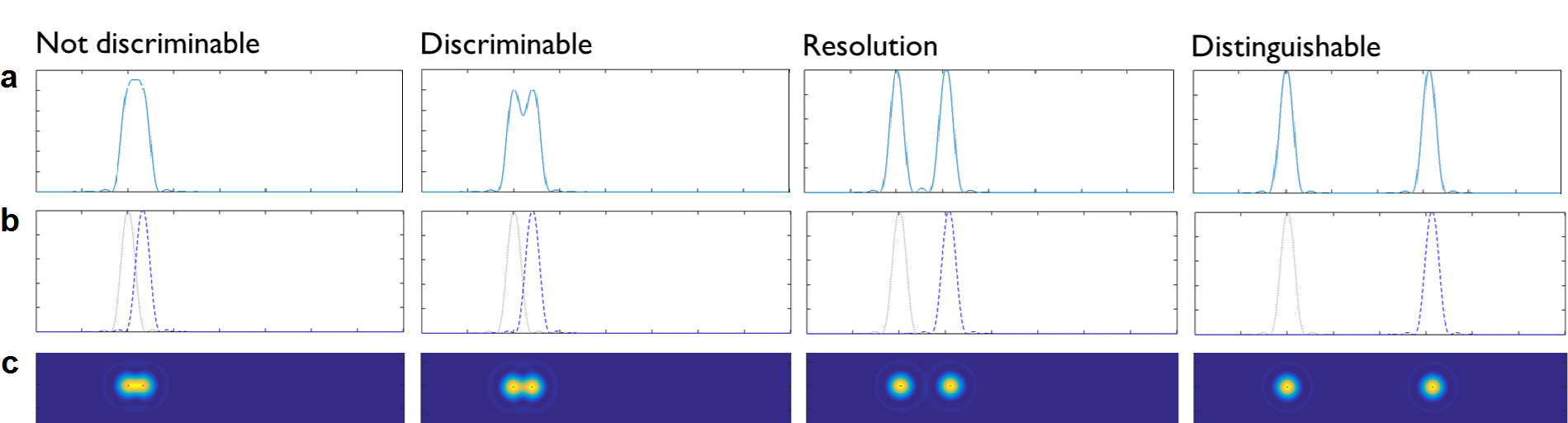}
\caption{Illustration of the concept of discriminability, resolution and distinguishability in optical microscopy. Total intensity profiles \textbf{(a)}, intensity profiles of each peak \textbf{(b)}, and focal plane views \textbf{(c)}.} 
\label{fig:resolution}
\end{figure}

The interpretation of the imaging of a single bead/organelle has relation to two fundamental concepts of the signal analysis~\cite{Urbanetal2015}: the resolution and the distinguishability. The microscopic resolution is defined as a distance at which two first-order valleys of the Airy waves~\cite{Airy1848} exactly merge. As seen in Fig.~\ref{fig:real_psf}, the Airy pattern is not observable in a real point (object) spread function. In contrast, the positions of the centroids of the interaction of light with a particle (i.e. intensity maximum/minimum) can be always found. Thus, searching for centroids of an object's response is a more realistic approach than the the theoretical Airy concept of resolution. Hence we suggested the information resolution concept~\cite{Rychtarikovaetal2016}.

The resolution can be also defined by absence or presence of points which do not have the identical intensity at the same position in the next z-level. In other words, a resolved object has to be surrounded by non-objects. Such an approach is critically dependent on the technical set-up and was discussed previously~\cite{Rychtarikovaetal2016}. 

The fluorescence imaging is interpreted much more simply than the diffractive imaging. Fluorescent molecules are of an infinitely small size (relative to a pixel) and we search for their distribution in a cell. Fig. \ref{fig:3D_fluorescent_cells}, \textit{upper} shows objects of unchanged intensities between two z-levels which were selected upon three simple assumptions:
\begin{enumerate} 
\item Each voxel contains one or none fluorophore which represents one intensity level.
\item When signals are present in numerous neighbouring pixels, these pixels can be assigned to an object. Unless signals in regions of lower density are of different intensity, they belong to the background. 
\item The voxels of different intensities are occupied by fluorophores of different spectral properties or there are more fluorophores in one voxel, i.e. they are relevant irrespective to the density of non-zero voxels in the neighbourhood. The detailed analysis showed that there are both (i) high intensity points in the centre of the dense areas -- the nucleoli -- which most likely represent several fluorophores per voxel and (ii) voxels of different intensities, mostly higher, outside dense regions, which most likely represent specific binding to objects other than nuclei, which is the primary target of the dye.
\end{enumerate}

If an image is full of labelled objects of different intensities (Fig.~\ref{fig:3D_fluorescent_cells}, \textit{lower}), it seems that any simple rule of data analysis cannot be applied. However, the assumption of zero point divergence gain significantly clarifies the dataset and enables a realistic 3D analysis of the observed structures. 

\section{Conclusion}

In this article we have highlighted a few important technical aspects which can limit the complete yield of information from a microscopic image:
\begin{enumerate} 
\item Usage of 12-bit colour depth of an image must not be sufficient, usage of a higher-bit depth can be necessary. 
\item It is legitimate to observe bright spots in difractive images. These spots are not necessarily (auto)fluorescent objects. 
\item The information can be localized with an infinite precision. It requests for microscopic cameras with a high number of pixels and usage of a higher magnification.
\end{enumerate}
Since samples are usually expensive and often irreplaceable, maximum of information from optical microscopic experiment should be acquired and analyzed.

\subsubsection*{Acknowledgments.} This work was supported by the Ministry of Education, Youth and Sports of the Czech Republic -- projects CENAKVA (No.\linebreak CZ.1.05/2.1.00/01.0024), CENAKVA II (No. LO1205 under the NPU I program), The CENAKVA Centre Development (No. CZ.1.05/2.1.00/19.0380) -- and by the CZ-A AKTION programme.

\end{document}